\documentclass[conference,a4paper]{IEEEtran}
\pdfoutput=1
\usepackage{amsmath,amssymb,mathrsfs}
\usepackage{epsfig,epsf,subfigure,graphicx,graphics}
\usepackage{url}

\newtheorem{lemma}{Lemma}[section]
\newtheorem{theorem}{Theorem}[section]

\newtheorem{definition}{Definition}[section]
\newtheorem{remark}{Remark}[section]

\newcommand{\E}{\mathrm{E}}

\graphicspath{{./figs/}}

\IEEEoverridecommandlockouts

\allowdisplaybreaks

\title{Coding with Encoding Uncertainty}
\author{
\authorblockN{Jad Hachem}
\authorblockA{UCLA, Los Angeles, USA\\
\textsf{jadhachem@ucla.edu}}
\and
\authorblockN{I-Hsiang Wang}
\authorblockA{
EPFL, Lausanne, Switzerland\\
\textsf{i-hsiang.wang@epfl.ch}}
\and
\authorblockN{Christina Fragouli}
\authorblockA{
EPFL, Lausanne, Switzerland\\
\textsf{christina.fragouli@epfl.ch}}
\and
\authorblockN{Suhas Diggavi}
\authorblockA{
UCLA, Los Angeles, USA\\
\textsf{suhasdiggavi@ucla.edu}}
\thanks{The work of J.\,Hachem and S.\,Diggavi was supported in part by NSF award 1136174 and MURI award AFOSR FA9550-09-064. The work of I.-H.\,Wang was supported by EU project CONECT FP7-ICT-2009-257616. The work of C.\,Fragouli was supported in part by the ERC Starting Grant Project NOWIRE ERC-2009-StG-240317 and the EU project STAMINA FP7-ICT-2009-265496.}
}

\renewcommand{\leadsto}{\to}
\DeclareMathOperator{\sgn}{sgn}
\DeclareMathOperator*{\argmin}{\arg\min}
\DeclareMathOperator*{\argmax}{\arg\max}
\newcommand{\pairwise}{P^{e(c\leadsto c')}}
\newcommand{\Lcal}{\mathcal{L}}
\newcommand{\Ccal}{\mathcal{C}}
\newcommand{\Bcal}{\mathcal{B}}
\newcommand{\Acal}{\mathcal{A}}
\newcommand{\Ical}{\mathcal{I}}
\usepackage{bbm}
\usepackage{enumerate}
\usepackage{comment}

\begin{document}
\maketitle
\begin{abstract}
We study the
channel coding problem when errors and uncertainty occur in the
encoding process. For simplicity we assume the channel between the
encoder and the decoder is perfect. Focusing on linear block codes, we
model the encoding uncertainty as erasures on the edges in the factor
graph of the encoder generator matrix. We first take a worst-case
approach and find the maximum tolerable number of erasures for perfect
error correction. Next, we take a probabilistic approach and derive a
sufficient condition on the rate of a set of codes, such that
decoding error probability vanishes as blocklength tends to
infinity. In both scenarios, due to the inherent asymmetry of the
problem, we derive the results from first principles, which indicates
that robustness to encoding errors requires new properties of codes
different from classical properties.
\end{abstract}

\section{Introduction}
In classical channel coding, the goal is to design encoding and decoding algorithms that combat errors and uncertainty introduced by the channel.
An implicit yet important assumption is that the encoder and decoder, once designed and implemented, operate in a deterministic and faultless manner. We ask in this paper, what if the encoder itself introduces uncertainty and errors?  

Encoder uncertainty can result from several causes. First, defects in a physical device that implements an encoder can make the encoding process faulty. Second, soft errors in processing and storage are becoming more frequent due to the trend in reducing the chip size. Third, as technology scales, variability in transistor design and device degradation also lead to unreliability. Lastly, errors can also happen in distributed encoding across physically separated devices which are connected through noisy channels (as in sensor networks).

The fact that we need to compute from unreliable components has been recognized in the literature,
but the focus has mainly been on computing (see, \emph{e.g.}, \cite{Pierce_65,Borkar_05}).
Recent work has also looked at the case where decoding might be subject to errors, in particular for message passing algorithms \cite{Varshney_11,Ruozzi_12}.
However, as far as we know, ours is the first work that tries to explore the effect of errors
during the encoding process.

As a first step in this direction, we focus in this paper on linear codes,
and model unreliability as \emph{erasures} on the \emph{edges} of the factor graph of the generator matrix.
As a result, an input message at the source will be mapped to one out of a set of codewords, with a certain probability.
Equivalently, we can think of the source as employing one out of a set of encoders, again with a certain probability.
The decoder needs to retrieve the message, knowing only the \emph{a priori} probability distribution on the encoders;
that is, the goal of the decoder is to recover the original codeword from the output of the defective encoder without knowing the realization of erasures.

Although this is, we believe, the simplest formulation, it is not a simple problem:
erasures on the edges of the factor graph lead to bit flips in the codewords, with a nonuniform probability,
that highly depends on the mapping from input messages to codewords, that is, on the structure of the generator matrix.
Trying to treat these errors as another source of noise, and incorporating them into the noisy channel, results in a channel model where the noise process has memory and depends on the generator matrix and input messages.  
Moreover, we found that even a small number of edge-erasures can significantly deteriorate the decoding performance. Hence, if such errors occur, it is necessary to design and employ codes that protect against them. We consider edge-erasures as a first step in this paper, but we can look at other error models as well (see Section~\ref{sec_furthererrormodels}).

In this paper, we focus for simplicity on the case where the channel
itself does not introduce errors, and make two contributions.  First,
we take a worst-case approach, in which we find the maximum number of
erasures that can occur, regardless of the erasure locations, such
that the decoder can always recover the original message. In other
words, we characterize the maximum number of erasures that can be
tolerated for perfect error correction. To this end, we define a new
distance metric between codewords to take the asymmetry of the
problem into account. Next, we take a probabilistic approach to analyze a minimum
distance decoder that emerges from the worst-case analysis. We assume that each edge is
erased with probability $p$, independently of other edges. We find a
sufficient condition on the rate of a code, under which decoding is
successful with vanishing error probability as the blocklength tends
to infinity. These results lead us to identify code properties that enable robustness to encoding
errors.

The rest of this paper is organized as follows. In Section~\ref{sec_Formulation}, we formulate the problem, and summarize the main results in Section~\ref{sec_Result}. The worst-case approach and a sketch of its analysis are presented in Section~\ref{sec_WorstCase}, while the probabilistic ones are given in Section~\ref{sec_Probabilistic}. The last section, Section~\ref{sec_furthererrormodels}, considers some additional error models.
The extended, detailed proofs are in the appendices: Appendix~\ref{app_WorstCase} gives the detailed proof of Section~\ref{sec_WorstCase}, while Appendix~\ref{app_Probabilistic} gives that of Section~\ref{sec_Probabilistic}. Finally, Appendix~\ref{app_Support} gives some mathematical results that support the proofs.

\section{Problem Formulation}\label{sec_Formulation}
We denote the generator matrix by $G\in\{0,1\}^{k\times n}$, with rate $R=\frac{k}{n}$. The message we wish to transmit is $m\in\{0,1\}^k$. The set of codewords associated with $G$ is $\Ccal(G)=\{c : c=mG\}$. The word received by the receiver is $r\in\{0,1\}^n$.

Erasures in the generator matrix occur as 1's in $G$ being flipped to 0's. Note that erasures do not always have an effect: for an erasure to affect a certain codeword bit, it must have occured on a bit in $G$ that will be multiplied by a 1 in $m$, not by a 0. This means that the error in the codeword depends on the message itself, and also on the generator matrix. Moreover, two erasures might cancel each other out, and hence different erasure patterns can give the same $r$.

We will present two error models, each of which will be used in one of the two approaches of this paper (worst-case and probabilistic).

\subsection{Error Models}\label{sec:formulation-errormodels}
\subsubsection{Worst-case Erasures}\label{sec:errormodel-worstcase}
In this model, erasures are introduced in $G$ that flip certain 1's in $G$
to 0's. We model this by the addition of what we call
an \emph{erasure matrix} $E\in\{0,1\}^{k\times n}$ to $G$. The
received word would then be $r=m(G+E)$ (binary addition) Note that $E$ has to satisfy a
certain condition: since the erasures only happen at 1's in $G$, $E$
can only have 1's at locations where $G$ also has 1's. In other words,
$G_{ij}=0$ implies $E_{ij}=0$. We denote this condition by $E\in2^G$
($E$ can be thought of as being 1 on a subset of the locations where
$G$ is 1, hence the power set notation). Moreover, we denote by
$2^G_{\eta}$ the subset of such $E$'s that have at most $\eta$
1's. Since each 1 denotes an erasure, $E\in2^G_\eta$ means there are
at most $\eta$ erasures. Effectively these errors alter the generator
matrix to $(G+E)$,  encoding $m$ as  $m(G+E)$ with $E$ unknown to both the encoder and the decoder.
The worst-case model allows any erasures
with these constraints.


\subsubsection{Probabilistic Erasures}\label{sec:errormodel-probabilistic}

In the probabilistic erasure model, we assume that each 1 in $G$ has a probability $p$ of being erased, \emph{i.e.}, of becoming a 0, independently of other 1's. We call these errors \emph{erasures}, and we denote their number by $\eta$, unless otherwise specified. However, it will be easier to deal with what we will refer to as \emph{bit errors}. These are the bit flips in the codewords themselves. Unless otherwise specified, we denote their number by $\eta'$.

\begin{definition}[Degree]
The probability of bit errors depends not only on the erasure probability $p$, but also on the following notion of `degrees': given a generator matrix $G$, we define the $\emph{degree}$ of each column $j$ as $d_j = \sum_{i=1}^k G_{ij}$, \emph{i.e.}, the number of 1's in column $j$ of $G$, for each $j\in\{1,\ldots,n\}$. We also define the same degree, relative to the message $m$, as $d_j^{(m)}=\sum_{i=1}^k m_iG_{ij}$. In other words, it is the number of 1's in column $G_j$ that, if flipped, will affect the outcome of the codeword bit $j$ when encoding message $m$.
\end{definition}

For a given message $m$, we can now compute the probability of a bit error. Bit $j$ will flip if there is an \emph{odd} number of erasures on the $d_j^{(m)}$ bits of column $G_j$ that affect the encoding of $m$. Since the number of erasures is a binomial random variable with probability $p$, the probability of a bit flip given a relative degree of $d_j^{(m)}=d$ is:
\begin{IEEEeqnarray}{rCcCl}
P_d &=& \sum_{\substack{0\le l\le d\\l\text{ is odd}}} \binom{d}{l}p^l(1-p)^{d-l} &=& \frac{1-\left(1-2p\right)^d}{2}
\end{IEEEeqnarray}

Note that, for $p\in\left(0,\frac{1}{2}\right)$, $P_d$ increases with $d$, and approaches $\frac{1}{2}$ from below as $d$ goes to infinity. Therefore, we can simplify things by letting $d^\ast=\max_jd_j$ (\emph{i.e.}, $d^\ast$ is the maximum degree of the columns of $G$) and by using the following bounds: for all $j\in\{1,\ldots,n\}$ and $m\in\{0,1\}^k$,
\begin{IEEEeqnarray*}{rCcCl?s}
P_{d_j^{(m)}}&\le& P_{d^\ast}&\\
P_{d_j^{(m)}} &\ge& P_1 &=& p & whenever $d_j^{(m)}\ge1$
\end{IEEEeqnarray*}

Finally, note that, given a message $m$ (or, equivalently, its corresponding codeword $c$), bit errors in each bit of the codeword are independent.
\begin{remark}
One may try to model such an encoder uncertainty as a noisy channel where the channel input is the original codeword and the output is the actual transmitted codeword by the defective encoder. Since the probability of bit errors on each symbol \emph{depends} on the message, that is, the whole input sequence, this channel has memory. 
More significantly, information bits are not protected by any channel code against such encoding uncertainty.
\end{remark}

Due to the asymmetry of the problem, the classical notion of codeword
distance is no longer useful. We will next define a new distance
metric, which is more relevant to this problem.

\subsection{New Notion of Distance}\label{sec:new-distance}

In classical problems, the (Hamming) distance $d_H$ between codewords
plays a critical role in code design and performance. It can be
interpreted as the minimum number of bits
($\left\lceil\frac{d_H}{2}\right\rceil$) required to `change' both
codewords into the same word.  In other words,
$\left\lceil\frac{d_H}{2}\right\rceil$ is the minimum number of bit
flips/errors on one codeword that would cause us to mistake it for the
other codeword.  For example, if we think of the classical Binary
Symmetric Channel, then $\left\lceil\frac{d_H}{2}\right\rceil$ is the
minimum number of errors that, if occurring on a codeword, would cause
us to not distinguish between it and another codeword:
$\left\lceil\frac{d_H}{2}\right\rceil$ errors could have happened on
either codeword and resulted in the same word. This last interpretation will be the most useful to adapt the concept of distance to our problem.

The biggest distinction to be made between our problem and classical problems is that the `changes' required between codewords are asymmetric. In the classical case, all bits can be flipped or erased, with no distinction. In our case, however, some bits can never flip.

For instance, take the zero codeword. Regardless of what the erasure matrix $E$ is, the zero codeword will always be encoded as $0\times(G+E)=0$. Hence, we can never `change' it into another one. Other codewords, however, can always be encoded as the zero codeword, for instance if $E=G$. So if we are comparing the zero codeword, $0$, with another codeword, $c$, of Hamming weight $d_H$, then, in order for $0$ and $c$ to be `changed' into the same word, we need only $\left\lceil\frac{d_H}{2}\right\rceil$ errors on each in the classical case, but we need $d_H$ bit flips on $c$ (and no bit flips on $0$) in our case for the same result.

To capture this phenomenon, we will use a ternary system for the codewords. Notice that, in the computation of a codeword $c=mG$, some zeros result from a sum of only 0's, and others result from a sum of 0's and an even (nonzero) number of 1's. The latter bits flip when an odd number of these 1's are erased. However, since there are no 1's in the sum of the former bits, they will \emph{never} flip.

This leads us to make a distinction between two types of zeros: those that may flip and become 1's, and those that never will. We call the former `soft zeros' and denote them by `$\bar0$', and call the latter `hard zeros' and denote them by `$0$'. Hence, all codewords $c$ are in the set $\{0,\bar0,1\}^n$. We will use this important $0$-$\bar0$ distinction to compute the new distance between any two codewords $c$ and $c'$, denoted by $\eta_0(c,c')$. As a natural extension to the classical distance, we define $\eta_0$ as the minimum number of erasures needed to encode both codewords as the same word.

To better describe $\eta_0(c,c')$, we will need to rearrange the bits of $c$ and $c'$ into nine categories, or blocks 
$\{\Bcal_i, i=1,\ldots,9\}$, 
as seen in Table~\ref{tbl:grouping-cc}.
\begin{table}[htbp]
\renewcommand{\arraystretch}{1.3}
\caption{Blocks $\Bcal_i$ grouping different pairs of bits of $c$ and $c'$. $w_i:=|\Bcal_i|$.}
\label{tbl:grouping-cc}
\centering
\begin{tabular}{|l|ccccccccc|}
\hline
block ($\Bcal_i$)&1&2&3&4&5&6&7&8&9\\
\hline
bits of $c$ &1&1&1&$\bar0$&$\bar0$&$\bar0$&0&0&0\\
bits of $c'$ &1&$\bar0$&0&1&$\bar0$&0&1&$\bar0$&0\\
\hline
\end{tabular}
\end{table}

Each $\Bcal_i$ contains all the bit locations where $c$ and $c'$ are as shown in column $i$ in the table. We denote their number by $w_i(c,c')$, or simply by $w_i$ when there is no ambiguity. So $\Bcal_4$, for example, contains all the bit locations for which $c$ is $\bar0$ and $c'$ is 1, and their number is $w_4$.

Now that we have formulated the problem, we can state the main results of the paper.

\section{Main Results}\label{sec_Result}
This paper approaches the problem from two different perspectives: a worst-case approach, and a probabilistic approach. The worst-case approach finds the maximum number of erasures that can occur, such that decoding is still perfect, \emph{i.e.,}\ it is always successful regardless of the erasure locations. The probabilistic approach finds the conditions under which decoding is successful with vanishing error probability (as the blocklength tends to infinity).

\subsection{Main Result: Worst-Case Approach}\label{sec:result-worstcase}

In the worst case approach, we assume the worst-case erasure error model described in Section~\ref{sec:errormodel-worstcase}. We use the following decoder:
\begin{quote}
Given a received word $r$, find the \emph{unique} codeword $c$ such that $c$ can be noisily encoded as $r$ for some erasure matrix $E$. If no such codeword exists, or more than one exist, then declare an error.
\end{quote}

For a given $G$, we wish to find the maximum number $\eta_{\max}$ such that, if no more than $\eta_{\max}$ erasures occur, then the above decoder will \emph{always} decode correctly, \emph{i.e.}, with \emph{zero error probability}. We call that: \emph{perfect decoding}.

\begin{theorem}\label{thm:eta0}
For a generator matrix $G$, the maximum number of erasures allowing perfect decoding is
\begin{IEEEeqnarray}{rCl}
\eta_{\max} &=& \min_{\substack{c,c'\in\Ccal(G)\\c\not=c'}} \eta_0(c,c') - 1
\end{IEEEeqnarray}
where $\eta_0(c,c')$ for any two codewords $c$ and $c'$ is the distance metric described in Section~\ref{sec:new-distance}, and equals:
\begin{IEEEeqnarray*}{rCl}
\label{eq:eta0}
\eta_0 &=& \left\{\,
\begin{IEEEeqnarraybox}[][c]{l?s}
\IEEEstrut
w_3 & if $w_3>w_7+w_2+w_4$\\
w_7 & if $w_7>w_3+w_2+w_4$\\
\left\lceil\frac{w_2+w_4+w_3+w_7}{2}\right\rceil & otherwise
\IEEEstrut
\end{IEEEeqnarraybox}\right.
\end{IEEEeqnarray*}
and the $w_i$'s are as defined in Table~\ref{tbl:grouping-cc}. Note that, for small $|w_3-w_7|$, we have $\eta_0=\left\lceil\frac{d_H}{2}\right\rceil$.
\end{theorem}

This result, to be proved in Section~\ref{sec_WorstCase}, can be interpreted similarly to that of a classical perfect
decoding problem. Indeed, to maximize the number of erasures allowed
for perfect decoding, we would want the (new notion of) distances
between codewords to be large: we
wish to increase the number of bits where two codewords are different
(\emph{i.e.,}\ blocks 2, 3, 4 and 7). However, there is more
importance on blocks 3 and 7 than on blocks 2 and 4. This is because
mistaking $c$ for $c'$ (resp., $c'$ for $c$) necessitates that (at
least) \emph{all} of the bits of $c$ (resp., $c'$) in block 3 (resp., block 7)
flip. Indeed, if $r$ has a $1$ in block 3, then we know that the
original codeword could not have been $c'$, since the latter is a hard $0$ in
that same block. On the other hand, if $r$ has a $1$ in block 2, then
$c'$ is still a possible codeword since it is a soft $\bar0$ in that block. Hence, $1$-$0$ bit differences are
more favorable than $1$-$\bar0$ bit differences.

\subsection{Main Result: Probabilistic Approach}\label{sec:result-probabilistic}

In the probabilistic approach, we assume the probabilistic error model described in Section~\ref{sec:errormodel-probabilistic}. We try to find the set of achievable rates $R$ for a given erasure probability $p$.
\begin{definition}[Achievable Rate]
A rate $R$ is called \emph{achievable} if there exists a set of encoding matrices $\left\{G_n\right\}_n$, $G_n\in\{0,1\}^{nR\times n}$, and corresponding decoders, such that the probability of error of these decoders, when encoding messages using $G_n$, goes to zero as $n$ goes to infinity.
\end{definition}

In this paper, we use the following minimum distance (MD) decoder to find an inner bound on the achievable rate:
\begin{quote}
Given a word $r$, find the codeword $c$ that minimizes the distance $\delta(c\to r)$.
\end{quote}
The distance $\delta$ is defined as $\delta(c\to r) = \sum_{i=1}^n \delta(c_i\to r_i)$, 
and $\delta$ for individual bits is explicitly defined in Table~\ref{tbl:delta}.
\begin{table}[htbp]
\renewcommand{\arraystretch}{1.3}
\caption{The distance metric $\delta$ on the bit level. Each cell represents $\delta(c_i\to r_i)$, where $c_i$ is in the top row and $r_i$ is in the left column.}
\label{tbl:delta}
\centering
\begin{tabular}{|c|ccc|}
\hline
$r_i\backslash c_i$ & 0 & $\bar0$ & 1\\
\hline
0 & 0 & 0 & 1 \\
1 & $\infty$ & 1 & 0 \\
\hline
\end{tabular}
\end{table}

Table~\ref{tbl:delta} can be explained as follows. The distance is infinite if $c$ can never be encoded as $r$ (\emph{i.e.}, if there is a 0 in $c$ that corresponds to a 1 in $r$). Otherwise, it is, just like with classical distance, the number of bit locations where $c$ and $r$ differ.

Just like $\eta_{\max}$ in Section~\ref{sec:result-worstcase} strongly depends on the $w_i$'s from Table~\ref{tbl:grouping-cc}, so will the conditions on the rate. However, since the $w_i$'s are linear functions of $n$, and we are taking a set of $G_n$'s with increasing $n$, we consider their normalized versions: $\alpha_i^{(n)}(c,c')=\frac{w_i(c,c')}{n}$. When there is no ambiguity, we simply denote these values by $\alpha_i$.

The new distance metric introduced and used in the worst-case approach
will also be useful in the probabilistic approach. However, it will be
more convenient to define a `directed' version of it. Specifically,
whereas $\eta_0(c,c')$ is the minimum number of erasures required to
mistake one of $c$ or $c'$ for the other, we define $\eta_0(c\leadsto
c')$ as the minimum number of erasures required to mistake $c$ for
$c'$, and \emph{not} the other way around. So the two quantities are
related by: $\eta_0(c,c') = \min\left\{\eta_0(c\leadsto
c'), \eta_0(c'\leadsto c)\right\}$

\begin{lemma}\label{lemma:eta0}
Given two codewords $c$ and $c'$, the minimum number of erasures required to mistake $c$ for $c'$, $\eta_0(c\leadsto c')$, is:
\begin{IEEEeqnarray*}{rCl}
\eta_0 &=& \left\{\,
\begin{IEEEeqnarraybox}[][c]{l?s}
\IEEEstrut
\infty & if $w_7>w_3+w_2+w_4$\\
w_3 & if $w_3>w_7+w_2+w_4$\\
\frac{w_2+w_4+w_3+w_7}{2} & otherwise
\IEEEstrut
\end{IEEEeqnarraybox}\right.
\end{IEEEeqnarray*}
\end{lemma}

The following theorem gives the sufficient condition on the rate. It will be useful, for the theorem as well as for its proof, presented in Section~\ref{sec_Probabilistic} and in Appendix~\ref{app_Probabilistic}, to introduce the following function $\psi$: for any number $x\ge0$, define $\psi(x)=x^x$.

\begin{theorem}\label{thm:rate}
Given a set of encoder matrices $\left\{G_n\right\}_n$ of rate $R$, a sufficient condition for a vanishing decoding error probability, using the MD decoder, is that, for all $n$:
\begin{IEEEeqnarray}{l'rCl}
\forall (c,c')\in\Ccal^2(G_n)\text{ s.t.\ } c\not=c',&R&<&-\log_2\beta(c\leadsto c')
\end{IEEEeqnarray}
where $\beta(c\leadsto c')$ is a term that depends on the two codewords $c$ and $c'$, with respect to the structure of $G_n$, specifically on the $\alpha_i$'s for this particular pair $(c,c')$. Equivalently, the sufficient condition can be rewritten as $R<-\log_2\beta_{\max}^{(n)}$, $\forall n$, where $\beta_{\max}^{(n)}$ is the maximum $\beta$ over all pairs $(c,c')$ in $\Ccal^2(G_n)$, with $c\not=c'$. For one such pair $(c,c')$, let $\alpha_0=\frac{\eta_0(c\leadsto c')}{n}$, $\gamma=(\alpha_2+\alpha_4+\alpha_1+\alpha_5)$, and $\alpha^\ast=\alpha_3+\frac{P_{d^\ast}}{P_{d^\ast}+1-p}\gamma$, and take $d^\ast$ to be the maximum degree of the columns of $G$. Then:
\begin{IEEEeqnarray}{rCl}
\beta &=& \left\{\,
\begin{IEEEeqnarraybox}[][c]{l?s}
\IEEEstrut
0 & if $\alpha_0=\infty$\\
P_{d^\ast}^{\alpha_3}\left(1-p\right)^{\alpha_6}\times(P_{d^\ast}+1-p)^\gamma & if $\alpha^\ast\ge\alpha_0$\\
P_{d^\ast}^{\alpha_3}\left(1-p\right)^{\alpha_6}\times \tilde\beta & otherwise
\IEEEstrut
\end{IEEEeqnarraybox}\right.\qquad
\end{IEEEeqnarray}
with
\begin{IEEEeqnarray*}{rCl}
\tilde\beta &=& P_{d^\ast}^{\frac{\alpha_2+ \alpha_4+\alpha_7-\alpha_3}{2}} (1-p)^{\alpha_1+\alpha_5+ \frac{\alpha_2+ \alpha_4-\alpha_7+\alpha_3}{2}}\\
&&\:\times\: \frac{\psi(\alpha_1+\alpha_5+\alpha_2+\alpha_4)}{\psi\left(\frac{\alpha_2+\alpha_4+\alpha_7-\alpha_3}{2}\right) \times\psi\left(\alpha_1+\alpha_5+ \frac{\alpha_2+ \alpha_4-\alpha_7+\alpha_3}{2}\right)}
\end{IEEEeqnarray*}
The $\tilde\beta$ term is a constant, $\tilde\beta<1$, that arises when $\alpha_0$ (equivalently, $\eta_0$) is large (but finite).
\end{theorem}

We now analyze $\beta(c\leadsto c')$ for a \emph{particular} (ordered)
pair $(c,c')$, and give an intuitive explanation of its
expression. Having large values for $\alpha_3$ and $\alpha_6$ will
result in a smaller $\beta$, which is favorable. This makes sense,
since, for codeword $c$ to be mistaken for codeword $c'$, all of the
bits of block 3 have to flip, and all of those of block 6 must stay
the same. (This intuitively explains the presence of
$P_{d^\ast}^{\alpha_3}\left(1-p\right)^{\alpha_6}$ in the expression
for $\beta$). Increasing $\alpha_3$ and $\alpha_6$ hence reduces the
probability that this happens. On the other hand, assuming these two
values are fixed, then having a large $\alpha_0$ is also favorable, as
it multiplies the whole expression by $\tilde\beta<1$, and divides it
by $(P_{d^\ast}+1-p)^\gamma>1$. This also makes sense, as a large
$\alpha_0$ means a larger number of bits required to mistake $c$ for
$c'$. Finally, an infinite $\alpha_0$, which is equivalent to `no
errors are possible', results in $\beta=0$, or
$R<\infty$, \emph{i.e.,} regardless of the rate, $c$ will not be
mistaken for $c'$.

However, reducing $\beta$ for one pair of codewords might increase it for another pair. Therefore, designing a code minimizing the maximum such $\beta$, for an overall good performance, is not straightforward, and is subject to further investigation.

\subsection{Discussion: Comparing the two approaches}

Both approaches indicate that having a large number of bits in blocks
3 and 7 is favorable (note that block 7 for the ordered pair $(c,c')$
is block 3 for $(c',c)$). Moreover, they both indicate that having
$1$-$0$ differences between codewords is more favorable than having $1$-$\bar0$
differences. However, the probabilistic approach also indicates that
block 6 (and, by symmetry, block 8) should be large. This means that
$\bar0$-$0$ differences are also important. It also takes into
consideration, in the expression of $\tilde\beta$, blocks 1 and 5,
where the bits are equal.

So, whereas the worst-case approach only considers which bits are
different, the probabilistic approach also looks at which bits are
equal. The results in Theorems \ref{thm:eta0} and \ref{thm:rate} give
guidelines for codes that are robust to worst-case and probabilistic
encoding errors respectively. This suggests that not all codes
designed for transmission errors may be robust to encoder errors. We
are currently investigating code designs based on these ideas.


\section{Worst-Case Approach}\label{sec_WorstCase}
Recall the error model described in Section~\ref{sec:errormodel-worstcase}, where the erasures can be thought of as a use of a different codebook than $\Ccal(G)$. Intuitively, perfect decoding can be ensured when these different codebooks do not contradict. To make this idea more rigorous, we introduce the notion of codebook ambiguity, which captures the nature of errors in the worst-case approach.

\subsection{Codebook Ambiguity}
\begin{definition}[Codebook Ambiguity]
We say that two generator matrices $G_1$ and $G_2$ have ambiguous codebooks, denoted by $\Acal(G_1,G_2)=1$, if $\exists m_1,m_2\in\{0,1\}^k$ such that $m_1\not=m_2$ and $m_1G_1=m_2G_2$.
\end{definition}

Since we will mostly be dealing with matrices of the form $(G+E)$, we write $\Acal^{(G)}(E_1,E_2)=\Acal(G+E_1,G+E_2)$.

Now we generalize this definition to all matrices $\{G+E:E\in2^G_\eta\}$ for some $\eta$. We use the following notation:
\begin{IEEEeqnarray*}{rCl}
\Acal^{(G)}(\le\eta) = \bigvee_{\substack{E_1,E_2\in2^G_\eta\\E_1\not=E_2}} \Acal^{(G)}(E_1,E_2)
\end{IEEEeqnarray*}
In other words, $\Acal^{(G)}(\le\eta)$ is 1 if and only if there are distinct $E_1,E_2\in2^G_\eta$ such that $(G+E_1)$ and $(G+E_2)$ have ambiguous codebooks.

The following lemma relates codebook ambiguity with perfect decoding:
\begin{lemma}\label{lemma:ambiguity}
Given an encoder matrix $G$, its $\eta_{\max}$, \emph{i.e.}, the maximum number of erasures such that perfect decoding is possible, is the largest integer such that $\Acal^{(G)}(\le\eta_{\max})=0$.
\end{lemma}

The lemma is useful as it gives us a method of computing $\eta_{\max}$, using codebook ambiguity. It follows directly from the definition of codebook ambiguity. Proof details are left for Appendix~\ref{app_WorstCase}.

\subsection{Proof of Theorem~\ref{thm:eta0} (sketch)}
We present a sketch of the proof, leaving the details for Appendix~\ref{app_WorstCase}.

The proof follows a simple procedure. We seek to find the minimum number of erasures $\eta$ such that $\Acal^{(G)}(\le\eta)=1$, \emph{i.e.}, such that there exist distinct $m_1$ and $m_2$, and distinct $E_1$ and $E_2$ with weight at most $\eta$, such that $m_1(G+E_1)=m_2(G+E_2)$. To do so, we go over all pairs of codewords $(c,c')$ (equivalently, pairs of messages $(m,m')$) and compute $\eta_0(c,c')$: the minimum $\eta$ such that $m(G+E_1)=m'(G+E_2)$ for some $E_1$, $E_2$ with weights at most $\eta$. Hence, $\Acal^{(G)}(\le\eta_0(c,c')) =1$ for all $(c,c')$, and $\Acal^{(G)}(\le\left(\min_{c,c'}\eta_0(c,c')-1\right))=0$. Therefore,
\begin{IEEEeqnarray}{rCl}
\eta_{\max} &=& \min_{\substack{c,c'\in\Ccal\\c\not=c'}} \eta_0(c,c') - 1
\label{eq:etamax}
\end{IEEEeqnarray}

To compute $\eta_0(c,c')$, we seek to count the minimum possible number of erasures that would confuse $c$ and $c'$. Specifically, we focus on bit errors on $c$ and $c'$. Since we only care about the \emph{minimum} number of erasures in $G$ that will cause a decoding error, each erasure has to correspond to exactly one bit error. So if $u$ is the number of bit errors on $c$, and $u'$ the number of bit errors on $c'$, such that $c$ and $c'$ both become the same word $r$, then $\eta_0$ will be the minimum value of $\max\{u,u'\}$ over all such $(u,u')$ pairs.

As we can see in Appendix~\ref{app_WorstCase}, following these steps will give us the value of $\eta_0$ expressed in Theorem~\ref{thm:eta0}.

\section{Probabilistic Approach}\label{sec_Probabilistic}
Assuming the probabilistic error model of Section~\ref{sec:errormodel-probabilistic}, we use the minimum distance (MD) decoder presented in Section~\ref{sec:result-probabilistic} to find an inner bound on the achievable rate. We will give a sketch of the analysis of the error probability of the MD decoder, which is itself the proof of the inner bound on the rate, described in Theorem~\ref{thm:rate}. The details of the analysis are given in Appendix~\ref{app_Probabilistic}.

\subsection{Analysis of the Probability of Error of the MD decoder (sketch)}

An error occurs when we encode a codeword $c$ into a word $r$, and decode $r$ as $c'$. We call this event a $c$-to-$c'$ error, denoted by $e(c\leadsto c')$.

The probability of a $c$-to-$c'$ error, \emph{i.e.}, the probability of decoding $c'$ given that we encoded $c$, is denoted by $\pairwise$. The average probability of error $P_e$ can be bounded by a function of $\pairwise$ as follows:
\begin{IEEEeqnarray}{rCl}
P_e &\le& 2^{nR} \max_{\substack{c,c'\\c\not=c'}}\pairwise
\label{eq:pe-general}
\end{IEEEeqnarray}
Note that the same bound applies for the maximal error probability (over all codewords $c$), since we bounded $P_e$ by the maximum error probability given $c$ and $c'$.

We will show that $\pairwise$ can be bounded by:
\begin{IEEEeqnarray}{rCl}
\label{eq:pairwise-1}
\pairwise &\le& n(n+1)C \times \left(\beta(c\leadsto c')\right)^n\notag\\
\max\pairwise &\le& n(n+1)C\left(\beta_{\max}^{(n)}\right)^n
\label{eq:pairwise}
\end{IEEEeqnarray}
where $C$ is a constant, and $\beta(c\leadsto c')$ depends only on the properties of the pair $(c,c')$. We also define $\beta_{\max}^{(n)}$ as the maximal $\beta$ over all pairs $(c,c')$ in $\Ccal^2(G_n)$, with $c\not=c'$. We can thus rewrite the bound on $P_e$ as:
\begin{IEEEeqnarray}{rCl}
P_e &\le& n(n+1)C\left(2^R\beta_{\max}^{(n)}\right)^n
\label{eq:pe}
\end{IEEEeqnarray}

This allows us to formulate the following sufficient condition for $P_e$ to decay to zero as $n\to\infty$:
\begin{IEEEeqnarray}{C?rCl?r}
\forall n:& R&<&-\log_2\beta(c\leadsto c'),&\forall(c,c')\text{ distinct}
\label{eq:inner-bound}
\end{IEEEeqnarray}
or, equivalently, $R<-\log_2\beta_{\max}^{(n)}$ for all $n$. Indeed, if \eqref{eq:inner-bound} is true, then $2^R\beta_{\max}^{(n)}<1$ and $P_e\to0$ as $n\to\infty$.

The rest of the proof, as shown in Appendix~\ref{app_Probabilistic}, consists of finding the expression of $\pairwise$ to deduce the expression of $\beta(c\leadsto c')$. We begin by providing a combinatorial expression for $\pairwise$ by conditioning over the number of bit errors $\eta'$, and averaging over that. Then, we use Stirling's approximation to provide an upper bound for $\pairwise$. Finally, we use some simple optimization techniques to bound the joint probability of the event ``$c$-to-$c'$ error and $\eta'$ bit errors'' by its maximum over $\eta'$. The resulting expression has the form in \eqref{eq:pairwise-1}, with the value of $\beta$ as given in Theorem~\ref{thm:rate}.

Following these steps proves Theorem~\ref{thm:rate}, giving an inner bound on the achievable rate of our problem.

\section{Further Error Models}\label{sec_furthererrormodels}
We discuss here some other potential error models. The most straightforward generalization of our edge-erasure error model is one where edges in the factor graph of $G$ are inserted as well as erased.

\subsection{Edge-Erasures and Edge-Insertions}
We have already described the edge-erasures in Section~\ref{sec_Formulation}. An edge-insertion error is one where a link between two nodes in the factor graph of the generator matrix $G$ is inserted with a certain probability. This corresponds to a 0 in $G$ becoming a 1. We can again regard these errors with both a worst-case approach, and a probabilistic approach.

\subsubsection{Worst-Case Approach}
In the worst-case approach, we try to find the maximum number of errors that can be tolerated for perfect decoding.

When we have both errors combined, the result is the addition of an erasure matrix $E\in\{0,1\}^{k\times n}$ to $G$ with \emph{no constraints}. The 1's of $E$ can be anywhere in the matrix.

This prompts us to undertake a similar analysis to the one in Section~\ref{sec_WorstCase}. It would be interesting to see if a distance metric similar to $\eta_0$ would also emerge. We have not yet explored this approach.

\subsubsection{Probabilistic Approach}
In the probabilistic approach, we want to find all rates such that the decoding error probability goes to zero as the blocklength tends to infinity. We show in this section that no such positive rate exists.

Indeed, as a natural extension of the edge-erasure error model, this error model can be characterized by two parameters: $p_0$ and $p_1$, where:
\begin{IEEEeqnarray*}{rCl}
p_0 &=& \Pr\left\{E_{ij}=1 | G_{ij}=0\right\}\\
p_1 &=& \Pr\left\{E_{ij}=1 | G_{ij}=1\right\}
\end{IEEEeqnarray*}
In other words, $p_0$ is the probability that a 0 in $G$ flips, and $p_1$ is the probability that a 1 in $G$ flips, where all these bit flips occur independently. For simplicity, we will consider, in the analysis that follows, the quantity $p=\min\{p_0,p_1\}$. We also assume that $p_0,p_1\in(0,\frac12)$.

We can now compute the probability of a bit flip in a codeword  $c$, given the codeword itself (or its corresponding information message $m$):
\begin{IEEEeqnarray*}{rCl}
\Pr\left\{r_j\not=c_j | m\right\}
&=& \Pr\left\{\left. \sum_{i=1}^{nR} m_i(G_{ij}+E_{ij}) \not= \sum_{i=1}^{nR} m_iG_{ij} \right| m \right\}\\
&=& \Pr\left\{\left. \sum_{i=1}^{nR} m_iE_{ij} = 1 \right| m \right\}\\
&=& \Pr\left\{\left. \sum_{i\in\Ical(m)} E_{ij} = 1 \right| m \right\}
\end{IEEEeqnarray*}
where $\Ical(m)$ is the set of indices $i$ such that $m_i=1$. The above can be reformulated as follows, using $\tilde E_i$ as independent Bernoulli random variables with parameter $p$:
\begin{IEEEeqnarray*}{rCl}
\Pr\left\{r_j\not=c_j | m\right\}
&\ge& \Pr\left\{\sum_{i=1}^{|\Ical(m)|} \tilde E_i = 1 \right\}\\
&=& \Pr\left\{\sum_{i=1}^{w(m)} \tilde E_i = 1 \right\}\\
&=& \frac{1-(1-2p)^{w(m)}}{2}
\end{IEEEeqnarray*}

For any real number $B$, the fraction of messages $m$ whose weight $w(m)$ becomes larger than $B$ approaches 1 as $n$ goes to infinity. This means that, asymptotically, almost all messages have a weight that approaches infinity. Therefore, the above probability approaches $\frac12$ almost surely as $n$ goes to infinity.

When the bit flips in a codeword occur with a probability of $\frac12$, then the resulting word is completely random, and hence the original message cannot be decoded. Therefore, for any $R>0$, no decoding scheme exists that asymptotically protects against both edge-erasures and edge-insertions.

\bibliographystyle{ieeetr}
\bibliography{Ref.bib}

\appendices

\section{Detailed Analysis of Worst-Case Approach}\label{app_WorstCase}
This appendix provides the details of the analysis of the worst-case approach. Specifically, it gives the proof of Lemma~\ref{lemma:ambiguity}, then computes the value of $\eta_0(c,c')$ for any two codewords $c$ and $c'$. These two results, combined with \eqref{eq:etamax}, imply the result of Theorem~\ref{thm:eta0}.

\subsection{Proof of Lemma~\ref{lemma:ambiguity}}
The proof of Lemma~\ref{lemma:ambiguity} is a simple application of the definition of codebook ambiguity. First, note that for any number of erasures $\eta\le\eta_{\max}$, perfect decoding is possible, and hence there are no two distinct messages $m_1$ and $m_2$ and no two distinct erasure matrices $E_1$ and $E_2$, of weight at most $\eta$, such that $m_1(G+E_1)=m_2(G+E_2)$. This implies that $\Acal^{(G)}(\le\eta_{\max})=0$. Conversely, if we have exactly $(\eta_{\max}+1)$ errors, then, by the definition of $\eta_{\max}$, there exist two distinct messages $m_1$ and $m_2$, and two distinct erasure matrices $E_1$ and $E_2$ with weight at most $(\eta_{\max}+1)$, such that $m_1(G+E_1)=m_2(G+E_2)$. Therefore, $\exists E_1,E_2\in2^G_{(\eta_{\max}+1)}$ distinct such that $\Acal^{(G)}(E_1,E_2)=1$, implying that $\Acal^{(G)}(\le(\eta_{\max}+1))=1$.

\subsection{Computation of $\eta_0(c,c')$}

The distance $\eta_0$ is equal to the minimum number of erasures required during the encoding of each codeword, so that both of them change to the same received word $r$.

The first observation to make is the following. Since what we observe is the codeword bit errors and not the erasures inside $G$, it makes sense to think of the former. Now note that a bit flips for an odd number of erasures in a relevant set of bits in $G$, and does not flip for an even number of those erasures. Since we are looking for the \emph{minimum} number of erasures, then one bit error will correspond to exactly one erasure, and a bit not in error will correspond to exactly zero erasures. Therefore, in the context of minimum number of erasures, erasures and bit errors are equivalent.

\subsubsection{Location of the bit errors}
The key point is to know the location of the bit errors that will cause the confusion between $c$ and $c'$. Recall that we want two things:
\begin{enumerate}[i.]
\item There must be enough errors to confuse $c$ and $c'$
\item We want the minimum number of such errors
\end{enumerate}

The goal is to change $c$ to $r$ and $c'$ to $r'$ so that $r_j=r_j'$ for every index $j$. We will think about that in terms of the nine blocks shown in Table~\ref{tbl:grouping-cc}.

In what follows, we assume $r=m(G+E_1)$ and $r'=m'(G+E_2)$, where $m$ and $m'$ are the messages corresponding to $c$ and $c'$, respectively, and $E_1$ and $E_2$ are erasure matrices. We define, for any erasure matrix $E$, its \emph{weight} $w(E)$ as the number of $1$'s in that matrix, \emph{i.e.}\ the number of erasures. If $E_1$ and $E_2$ cause confusion between $c$ and $c'$, then the number of erasures required for that is $\max\{w(E_1), w(E_2)\}$. Hence, $\eta_0(c,c')$ is the minimum such value over all valid erasure matrices $E_1$ and $E_2$:
\begin{IEEEeqnarray*}{rCl}
\eta_0(c,c') &=& \min_{\substack{E_1,E_2\in2^G\\m(G+E_1)=m'(G+E_2)}} \max\{w(E_1),w(E_2)\}
\end{IEEEeqnarray*}

To have the minimum number of errors, we should not flip bits that are already equal to each other. Hence, blocks 1, 5, 6, 8 and 9 should not contain any errors.

In blocks 3 and 7, errors can only arise on one of the two codewords, namely, the one that contains the $1$'s. All of the bits in the blocks have to become equal, so all of the $1$'s have to flip. This means $w_3$ errors in $E_1$, and $w_7$ errors in $E_2$.

This leaves us with blocks 2 and 4. Here, we could have errors on both codewords. All bits have to be flipped on exactly one of the two codewords. Let $\eta_1$ denote the number of bits flipped in block 4 on codeword $c$ (so the remaining $(w_4-\eta_1)$ bits are flipped on $c'$), and $\eta_1'$ denote the number of bits flipped in block 2 on codeword $c'$ (the remaining $(w_2-\eta_1')$ are flipped on $c$). This adds another $\eta_1+(w_2-\eta_1')$ errors in $E_1$, and $\eta_1'+(w_4-\eta_1)$ errors in $E_2$.

Summing all of these errors, we get:
\begin{IEEEeqnarray}{l"r}
\qquad&\left\{\,
\begin{IEEEeqnarraybox}[][c]{rCl}
\IEEEstrut
w(E_1) &=& w_3 + \eta_1 + (w_2-\eta_1')\\
w(E_2) &=& w_7 + \eta_1' + (w_4-\eta_1)
\IEEEstrut
\end{IEEEeqnarraybox}\right.\notag\\
\IEEEeqnarraymulticol{2}{s}{and therefore,}\notag\\
\label{eq:error-matrices-weights}
&\left\{\,
\begin{IEEEeqnarraybox}[][c]{rCl}
\IEEEstrut
w(E_1) &=& w_3+w_2 + (\eta_1 - \eta_1')\\
w(E_2) &=& w_7+w_4 - (\eta_1 - \eta_1')
\IEEEstrut
\end{IEEEeqnarraybox}\right.
\end{IEEEeqnarray}

\subsubsection{Optimizing over $(\eta_1-\eta_1')$}
The minimal number of errors that could cause confusion between codewords $c$ and $c'$ necessitates $E_1$ and $E_2$ to satisfy the conditions in \eqref{eq:error-matrices-weights}. The conditions depend on a parameter $\Delta\eta=(\eta_1-\eta_1')\in\left\{-w_2,\ldots,w_4\right\}$. Hence, we get:
\begin{IEEEeqnarray*}{rl}
\eta_0(c,c') = \min_{\Delta\eta\in\left\{-w_2,\ldots,w_4\right\}}\ \max&\left\{w_3+w_2+\Delta\eta,\right.\\
&\left.\:\:w_7+w_4-\Delta\eta\right\}
\end{IEEEeqnarray*}

Optimizing over $\Delta\eta$, we get the following three cases:
(recall $-w_2\le\Delta\eta\le w_4$)

\paragraph{Case 1: $w_3\ge w_7+w_2+w_4$}
In this case, recall that, for all $\Delta\eta\in\{-w_2,\ldots,w_4\}$,
\begin{IEEEeqnarray*}{s?rCl}
&w_3+(w_2+\Delta\eta) &\ge& w_3\\
\IEEEeqnarraymulticol{4}{s}{and}\\
&w_7+w_2+w_4 &\ge& w_7+w_2+w_4-(w_2+\Delta\eta)\\
&&\ge& w_7+w_4-\Delta\eta\\
\IEEEeqnarraymulticol{4}{s}{and hence,}\\
&w_3+(w_2+\Delta\eta) &\ge& w_7+w_4-\Delta\eta\\
\IEEEeqnarraymulticol{4}{s}{Therefore,}\\
&\eta_0 &=& \min_{\Delta\eta} \left(w_3+w_2+\Delta\eta\right)\\
&&=& \left.\left(w_3+w_2+\Delta\eta\right)\right|_{\Delta\eta=-w_2}\\
&&=& w_3
\end{IEEEeqnarray*}

So in this case, $\eta_0=w_3$.

\paragraph{Case 2: $w_7\ge w_3+w_2+w_4$}
Using a similar argument, we can show that, in this case, $\eta_0=w_7$.

\paragraph{Case 3: both conditions are false}
One can think of $(w_3+w_2+\Delta\eta)$ and $(w_7+w_4-\Delta\eta)$ as two lines in the plane, as a function of $\Delta\eta$: one with slope $(+1)$ and the other with slope $(-1)$. Their maximum is therefore a V-shaped function whose minimum is the intersection of the two lines. Hence, the $\Delta\eta$ that minimizes the maximum of the two functions---call it $\Delta\eta^\ast$---is such that
\begin{IEEEeqnarray*}{s?rCl}
&w_3+w_2+\Delta\eta^\ast &=& w_7+w_4-\Delta\eta^\ast\\
&\Delta\eta^\ast &=& \frac{1}{2}\left(w_7-w_3+w_4-w_2\right)\\
\IEEEeqnarraymulticol{4}{s}{and $\Delta\eta^\ast\in[-w_2,w_4]$ since we are in case 3. Therefore,}\\
&\eta_0 &=& w_3+w_2+\Delta\eta^\ast\\
&&=& \frac{w_2+w_4+w_3+w_7}{2}
\end{IEEEeqnarray*}

However, since $\eta_0$ can only be an integer, we take the ceiling of that last value. Therefore, in this case, we have $\eta_0=\left\lceil \frac{w_2+w_4+w_3+w_7}{2}\right\rceil$.

This concludes the proof of Theorem~\ref{thm:eta0}.

\section{Detailed Analysis of Probabilistic Approach}\label{app_Probabilistic}
This appendix provides details on the analysis of the probabilistic approach. It deals with proving Theorem~\ref{thm:rate} by elaborating on the proof sketch given in Section~\ref{sec_Probabilistic}.

\subsection{Analysis of the Probability of Error of the MD decoder}

Recall that $\pairwise$ denotes the probability of decoding $c'$ given that we encoded $c$. A bound on the average probability of error $P_e$ can be derived as a function of $\pairwise$ as follows:
\begin{IEEEeqnarray}{rCl}
P_e &=& \Pr\{\exists c,c'\text{ distinct s.t.\ we have a $c$ to $c'$ error}\}\notag\\
&\le& \sum_c\sum_{c'\not=c} \Pr\{\text{encoding $c$}\}\pairwise\notag\\
&\le& \sum_c\sum_{c'\not=c} 2^{-nR}\pairwise\notag\\
&\le& 2^{nR}\left(2^{nR}-1\right)2^{-nR} \max_{\substack{c,c'\\c\not=c'}}\pairwise\notag\\
&\le& 2^{nR} \max_{\substack{c,c'\\c\not=c'}}\pairwise
\label{eq:pe-general}
\end{IEEEeqnarray}

Note that the same bound applies for the maximal error probability (over all codewords $c$), since we bounded $P_e$ by the maximum error probability given $c$ and $c'$.

We will show that $\pairwise$ can be bounded by:
\begin{IEEEeqnarray}{rCl}
\pairwise &\le& n(n+1)C\times \left(\beta(c\leadsto c')\right)^n\notag\\
\max\pairwise &\le& n(n+1)C\left(\beta_{\max}^{(n)}\right)^n
\label{eq:pairwise}
\end{IEEEeqnarray}
where $C$ is a constant, and $\beta(c\leadsto c')$ depends only on the properties of the pair $(c,c')$. We also define $\beta_{\max}^{(n)}$ as the maximal $\beta$ over all pairs $(c,c')\in\Ccal^2(G_n)$, with $c\not=c'$. We can thus rewrite the bound on $P_e$ as:
\begin{IEEEeqnarray}{rCl}
P_e &\le& 2^{nR} \times n(n+1)C \left(\beta_{\max}^{(n)}\right)^n\notag\\
P_e &\le& n(n+1)C\left(2^R\beta_{\max}^{(n)}\right)^n
\label{eq:pe}
\end{IEEEeqnarray}

This allows us to formulate the following sufficient condition for $P_e$ to decay to zero as $n\to\infty$:
\begin{IEEEeqnarray}{rCl}
R &<& -\log_2\beta_{\max}^{(n)}
\label{eq:inner-bound}
\end{IEEEeqnarray}

Indeed, if \eqref{eq:inner-bound} is true, then $2^R\beta_{\max}^{(n)}<1$ and $P_e\to0$ as $n\to\infty$.

In the following, we will compute the value of $\beta_{\max}^{(n)}$ for both decoders. Recall the bit rearrangements in Table \ref{tbl:grouping-cc}: each block $\Bcal_i$ has a size $w_i$, normalized as $\alpha_i=\frac{w_i}{n}$.

\subsubsection{Computing the pairwise error probability $\pairwise$}
Given that we are encoding codeword $c$, we want to find the probability of mistaking it for codeword $c'$. We can write it as follows:
\begin{IEEEeqnarray}{rCl}
\pairwise &=& \sum_{\eta'=0}^n \pairwise(\eta')\notag\\
&\le& (n+1)\max_{\eta'} \pairwise(\eta')
\end{IEEEeqnarray}
where $\pairwise(\eta')$ is the joint probability of having a $c$-to-$c'$ error, \emph{and} having $\eta'$ bit errors. We can write it as
\begin{IEEEeqnarray}{rCl}
\pairwise(\eta') &=& P(\eta')\pairwise_{|\eta'}
\end{IEEEeqnarray}
where now $\pairwise_{|\eta'}$ is the probability of having a $c$-to-$c'$ error \emph{given} that we have $\eta'$ bit errors, and $P(\eta')$ is the probability of having $\eta'$ bit errors.

To compute $P(\eta')$, first note that bit errors in $c$ can only occur on blocks $\Bcal_1$ through $\Bcal_6$. Each bit in these blocks has a probability at most $P_{d^\ast}$ of flipping, and a probability at most $(1-p)$ of not flipping, independently of the other bits. Therefore,
\begin{IEEEeqnarray}{rCl}
P(\eta') &\le& \binom{w_1+\cdots+w_6}{\eta'}P_{d^\ast}^{\eta'}\left(1-p\right)^{w_1+\cdots+w_6-\eta'}\notag\\
&\le& \binom{\bar w}{\eta'}P_{d^\ast}^{\eta'}\left(1-p\right)^{\bar w-\eta'}
\end{IEEEeqnarray}
where $\bar w=w_1+\cdots+w_6$.

To compute $\pairwise_{|\eta'}$, we have to revisit the distance $\eta_0(c\leadsto c')$ introduced in Section \ref{sec:result-probabilistic}.

Recall the statement of Lemma~\ref{lemma:eta0}, which says that, given two codewords $c$ and $c'$, the minimum number of erasures $\eta_0(c\leadsto c')$ required to mistake $c$ for $c'$ is
\begin{IEEEeqnarray*}{rCl}
\eta_0 &=& \left\{\,
\begin{IEEEeqnarraybox}[][c]{l?s}
\IEEEstrut
\infty & if $w_7>w_3+w_2+w_4$\\
w_3 & if $w_3>w_7+w_2+w_4$\\
\frac{w_2+w_3+w_4+w_7}{2} & otherwise
\IEEEstrut
\end{IEEEeqnarraybox}\right.
\end{IEEEeqnarray*}
Specifically, these errors occur in the following way: all of the bits in block 3 must flip, and the remaining bit errors have to occur somewhere in blocks 2 and 4.

\paragraph*{Proof of Lemma~\ref{lemma:eta0}}
Assume $\eta$ bit errors have occurred. To confuse $c$ for $c'$, these errors must `change' $c$ to a word $r$ such that $\delta(c\to r)>\delta(c'\to r)$. First note that the bits of block 6 must not flip, otherwise $r$ cannot result from $c'$ (and hence $\delta(c'\to r)=\infty$). Second, bit flips in blocks 1 and 5 affect both distances the same way, so we ignore them as they play no role in confusing $c$ for $c'$. Third, all $w_3$ bits in block 3 must flip. The remaining $\eta-w_3$ bit flips have to occur in blocks 2 and 4. The distance between $c$ and $r$ is the number of bit flips, $\eta$, whereas the distance between $c'$ and $r$ is the number of bits in block 7, plus the number of bits of blocks 2 and 4 that did not flip. Hence,
\begin{IEEEeqnarray*}{RcCl}
&\delta(c\to r)&>&\delta(c'\to r)\\
\iff& \eta &>& w_7 + \bigl(w_2+w_4-(\eta-w_3)\bigr)\\
\iff& \eta &>& \frac{w_3+w_7+w_2+w_4}{2} = \tilde\eta_0
\end{IEEEeqnarray*}
But $\eta$ cannot be larger than $w_3+w_2+w_4$. So if $\tilde\eta_0>w_3+w_2+w_4$, then we will never have $\eta>\tilde\eta_0$, and so the distance from $c$ to $c'$, $\eta_0$, is infinite. Otherwise, since we have to have at least $w_3$ bit errors for a $c$-to-$c'$ error in the first place, then $\eta_0=\max\{w_3,\lceil\tilde\eta_0\rceil\}$. This completes the proof of Lemma~\ref{lemma:eta0}.
\begin{remark}
The definitions of $\eta_0(\cdot,\cdot)$ and $\eta_0(\cdot\to\cdot)$ agree with the statement:
\begin{IEEEeqnarray*}{rCl}
\eta_0(c,c') &=& \min\left\{ \eta_0(c\to c') , \eta_0(c'\to c) \right\}
\end{IEEEeqnarray*}
\end{remark}

Given that we have exactly $\eta'$ errors, they can be anywhere in blocks 1 through 6. So the total number of possible error distributions is $\binom{\bar w}{\eta'}$. However, the bit error distributions that actually cause us to mistake $c$ for $c'$ have to satisfy the following conditions:
\begin{itemize}
\item $w_3$ of these errors have to occur in block 3
\item $l\ge\eta_0-w_3$ errors have to occur in blocks 2 and 4
\item The remaining $\eta'-l-w_3$ errors must be in blocks 1 and 5 (\emph{not} in block 6)
\end{itemize}

The first two conditions ensure that we have at least $\eta_0$ errors in the minimum locations required for a $c$-to-$c'$ error. The third condition ensures that there are no errors in block 6, because otherwise, a bit in block 6 would be 1, and we would know that the original codeword could not have been $c'$, which takes as value a soft 0 in block 6.

So the total number of bit error distributions that cause a $c$-to-$c'$ error is
\begin{IEEEeqnarray}{rCl}
\binom{w_3}{w_3}\sum_{l=\eta_0-w_3}^{w_2+w_4}\binom{w_2+w_4}{l}\binom{w_1+w_5}{\eta'-(l+w_3)}
\end{IEEEeqnarray}
and so
\begin{IEEEeqnarray}{rCl}
\pairwise_{|\eta'} &=& \frac{\sum_{l=\eta_0-w_3}^{w_2+w_4}\binom{w_2+w_4}{l}\binom{w_1+w_5}{\eta'-(l+w_3)}}{\binom{\bar w}{\eta'}}
\end{IEEEeqnarray}
which results in
\begin{IEEEeqnarray*}{rCl}
\IEEEeqnarraymulticol{3}{l}{\pairwise(\eta')}\\
\quad&\le& P_{d^\ast}^{\eta'}\left(1-p\right)^{\bar w-\eta'}\\
&&\:\times \sum_{l=\eta_0-w_3}^{w_2+w_4}\binom{w_2+w_4}{l}\binom{w_1+w_5}{\eta'-(l+w_3)}\\
&\le& \left[P_{d^\ast}^{\alpha'}\left(1-p\right)^{\bar\alpha-\alpha'}\right]^n\\
&&\:\times \sum_{l=\eta_0-w_3}^{w_2+w_4}\binom{(\alpha_2+\alpha_4)n}{\lambda n}\binom{(\alpha_1+\alpha_5)n}{(\alpha'-\alpha_3-\lambda)n} \quad\\
&\le& \left[P_{d^\ast}^{\alpha'}\left(1-p\right)^{\bar\alpha-\alpha'}\right]^n\\
&&\:\times\: n\times\max_{\lambda\in\Lcal} \binom{(\alpha_2+\alpha_4)n}{\lambda n}\binom{(\alpha_1+\alpha_5)n}{(\alpha'-\alpha_3-\lambda)n} \quad \IEEEyesnumber
\label{eq:pairwise-combinatorial}
\end{IEEEeqnarray*}
where $\alpha'=\frac{\eta'}{n}$ and $\lambda=\frac{l}{n}$. We will also use $\alpha_0=\frac{\eta_0}{n}$ later on. $\Lcal$ is defined as the set of $\lambda$'s that satisfy:
\begin{IEEEeqnarray*}{rCcCl}
0\le\alpha_0-\alpha_3&\le&\lambda&\le&\alpha_2+\alpha_4\\
(\alpha'-\alpha_3)-(\alpha_1+\alpha_5)&\le&\lambda&\le&\alpha'-\alpha_3
\end{IEEEeqnarray*}

Stirling's approximation gives, for any integer $m\ge1$:
\begin{IEEEeqnarray}{c}
\sqrt{2\pi}\times \sqrt{m}\left(\frac{m}{e}\right)^m \le m! \le e\times \sqrt{m}\left(\frac{m}{e}\right)^m
\end{IEEEeqnarray}
(also, the left inequality would trivially hold for $m=0$)

Using this in binomial coefficients, we get, for all integers $m\ge1$ and $0\le k\le m$:
\begin{IEEEeqnarray}{rCl}
\binom{m}{k} &\le& \frac{m!}{k!(m-k)!}\notag\\
&\le& \frac{e\sqrt{m}\left(\frac{m}{e}\right)^m}{\sqrt{2\pi k}\left(\frac{k}{e}\right)^k\sqrt{2\pi (m-k)}\left(\frac{m-k}{e}\right)^{m-k}}\notag\\
&\le& \frac{e}{2\pi} \sqrt{\frac{m}{k(m-k)}} \frac{m^m}{k^k(m-k)^{m-k}}
\end{IEEEeqnarray}

For $k=0$ or $k=m$, we have $\binom{m}{k}=1$. Since the above bound gives infinity in both these cases, we can replace it with:
\begin{IEEEeqnarray}{s?rCl}
&\binom{m}{k} &\le& \frac{e}{2\pi} \sqrt{A_{m,k}} \frac{m^m}{k^k(m-k)^{m-k}}\\
where& A_{m,k} &=& \left\{\,
\begin{IEEEeqnarraybox}[][c]{l?s}
\IEEEstrut
1 & if $k=0$ or $k=m$\\
\frac{m}{k(m-k)} & otherwise
\IEEEstrut
\end{IEEEeqnarraybox}\right.
\end{IEEEeqnarray}

We show, in Appendix~\ref{app:binom-bound}, that $A_{m,k}\le2$ for all integers $m\ge1$ and $0\le k\le m$. Therefore, we can write:
\begin{IEEEeqnarray}{rCl}
\binom{m}{k} &\le& \frac{e}{2\pi} \sqrt{2} \frac{m^m}{k^k(m-k)^{m-k}}\notag\\
&\le& \frac{e}{\pi\sqrt{2}} \frac{m^m}{k^k(m-k)^{m-k}}
\label{eq:binom-stirling}
\end{IEEEeqnarray}

Since everything is normalized by $n$, we can write \eqref{eq:binom-stirling}, for some $0<r\le1$ and $0\le l\le r$, such that $rn$ and $ln$ are integers, as:
\begin{IEEEeqnarray}{rCl}
\binom{rn}{ln} &\le& \frac{e}{\pi\sqrt{2}} \frac{(rn)^{rn}}{(ln)^{ln}(rn-ln)^{rn-ln}}\notag\\
&\le& \frac{e}{\pi\sqrt{2}} \left(\frac{r^r}{l^l(r-l)^{r-l}}\right)^n
\end{IEEEeqnarray}

Using this in \eqref{eq:pairwise-combinatorial}:
\begin{IEEEeqnarray}{rCl}
\pairwise(\eta') &\le& nC\left[P_{d^\ast}^{\alpha'}\left(1-p\right)^{\bar\alpha-\alpha'}\right]^n \max_{\lambda\in\Lcal}\ b(\lambda)^n\quad
\end{IEEEeqnarray}
where $C=\left(\frac{e}{\pi\sqrt{2}}\right)^2$ is a constant. To express $b(\lambda)$, we recall the function $\psi(x)=x^x$, $\forall x\ge0$, and get:
\begin{IEEEeqnarray}{rCl}
b(\lambda) &=& \frac{\psi(\alpha_2+\alpha_4)}{\psi(\lambda) \psi(\alpha_2+\alpha_4-\lambda)}\IEEEnonumber\\
&&\:\times\:\frac{\psi(\alpha_1+\alpha_5)}{\psi(\alpha'-\alpha_3-\lambda) \psi(\alpha_1+\alpha_5+\alpha_3-\alpha'+\lambda)}\quad
\end{IEEEeqnarray}

By noting that $x^n$ is an increasing function, we can formulate the bound on $\pairwise$ as:
\begin{IEEEeqnarray*}{rCl}
\IEEEeqnarraymulticol{3}{l}{\pairwise}\\
\quad&\le& n(n+1)C \left[ \max_{\alpha_0\le\alpha'\le\bar\alpha}\left( P_{d^\ast}^{\alpha'}\left(1-p\right)^{\bar\alpha-\alpha'} \max_{\lambda\in\Lcal}\ b(\lambda)\right) \right]^n\\
&\le& n(n+1)C \left[ \max_{\alpha_0\le\alpha'\le\bar\alpha}\left( P_{d^\ast}^{\alpha'}\left(1-p\right)^{\bar\alpha-\alpha'} b^\ast \right) \right]^n\\
&\le& n(n+1)C \left[ \max_{\alpha_0\le\alpha'\le\bar\alpha} \beta(\alpha') \right]^n\\
&\le& n(n+1)C \Bigl[ \beta(c\leadsto c') \Bigr]^n\IEEEyesnumber
\label{eq:pairwise-beta}
\end{IEEEeqnarray*}

By substituting this in \eqref{eq:pe-general}, it follows that
\begin{IEEEeqnarray*}{rCl}
P_e &\le& n(n+1)C\Bigl[2^R\beta_{\max}^{(n)}\Bigr]^n
\end{IEEEeqnarray*}
where $\beta_{\max}^{(n)}=\max_{c\not=c'\in\Ccal(G_n)}\beta(c\leadsto c')$.

\subsubsection{Proof of Theorem~\ref{thm:rate}}
To prove Theorem~\ref{thm:rate}, we have to compute the value of $\beta(c\leadsto c')$. This is given in Lemma~\ref{lemma:beta} below. 

\begin{lemma}\label{lemma:beta}
Given an ordered pair of codewords: $(c,c')$, define $\gamma=(\alpha_2+\alpha_4+\alpha_1+\alpha_5)$, and $\alpha^\ast=\alpha_3+P_{d^\ast}\gamma$. Then, the $\beta(c\leadsto c')$ term can be expressed as
\begin{IEEEeqnarray*}{rCl}
\beta(c\leadsto c') &=& \left\{\,
\begin{IEEEeqnarraybox}[][c]{l?s}
\IEEEstrut
0 & if $\alpha_0=\infty$\\
P_{d^\ast}^{\alpha_3}\left(1-p\right)^{\alpha_6}\times (P_{d^\ast}+1-p)^\gamma & if $\alpha^\ast\ge\alpha_0$\\
P_{d^\ast}^{\alpha_3}\left(1-p\right)^{\alpha_6}\times \tilde\beta & otherwise
\IEEEstrut
\end{IEEEeqnarraybox}\right.\qquad
\end{IEEEeqnarray*}
where
\begin{IEEEeqnarray*}{rCl}
\tilde\beta &=& P_{d^\ast}^{\frac{\alpha_2+ \alpha_4+\alpha_7-\alpha_3}{2}} (1-p)^{\alpha_1+\alpha_5+ \frac{\alpha_2+ \alpha_4-\alpha_7+\alpha_3}{2}}\\
&&\:\times\: \frac{(\alpha_1+\alpha_5+\alpha_2+\alpha_4) }{\left(\frac{\alpha_2+\alpha_4+\alpha_7-\alpha_3}{2}\right)\left(\alpha_1+\alpha_5+ \frac{\alpha_2+ \alpha_4-\alpha_7+\alpha_3}{2}\right)}\\
&<& 1
\end{IEEEeqnarray*}
\end{lemma}

Note that Theorem~\ref{thm:rate} follows directly from Lemma~\ref{lemma:beta}. Indeed, if $R<-\log_2\beta(c\leadsto c')$ for all ordered pairs of distinct codewords $(c,c')$, or, equivalently, if $R<-\log_2\beta_{\max}^{(n)}$, then $P_e\to0$ as $n\to\infty$. Substituting $\beta$ with the value shown in Lemma~\ref{lemma:beta} gives the exact statement of Theorem~\ref{thm:rate}. Therefore, all that is left to do is to prove Lemma~\ref{lemma:beta}.

\subsubsection{Proof of Lemma~\ref{lemma:beta}}
First, recall that, if $\alpha_0=\infty$, then the pairwise error probability $\pairwise$ is zero, and so we can write $\beta(c\leadsto c')=0$ in that case. So we now assume that $\alpha_0$ is finite.

As seen in \eqref{eq:pairwise-beta}, we have:
\begin{IEEEeqnarray*}{rCl}
\beta(c\leadsto c') &=& \max_{\alpha'\in\left[\alpha_0,\bar\alpha\right]} \beta(\alpha')\\
\beta(\alpha') &=& P_{d^\ast}^{\alpha'}(1-p)^{\bar\alpha-\alpha'}b^\ast\\
b^\ast &=& \max_{\lambda\in\Lcal} b(\lambda)
\end{IEEEeqnarray*}

We will proceed with computing $b^\ast$, then $\beta(c\leadsto c')$, which we will denote by $\beta^\ast$ for simplicity in what follows.

\paragraph{Computing $b^\ast$}
Recall the expression of $b(\lambda)$:
\begin{IEEEeqnarray*}{rCl}
b(\lambda) &=& \frac{\psi(\alpha_2+\alpha_4)}{\psi(\lambda) \psi(\alpha_2+\alpha_4-\lambda)}\IEEEnonumber\\
&&\:\times\:\frac{\psi(\alpha_1+\alpha_5)}{\psi(\alpha'-\alpha_3-\lambda) \psi(\alpha_1+\alpha_5+\alpha_3-\alpha'+\lambda)}\quad
\end{IEEEeqnarray*}

As can be seen in Appendix~\ref{app:derivations}, we can write:
\begin{IEEEeqnarray*}{rCl}
\sgn\{b'(\lambda)\} &=& \sgn\left\{ \log\frac{(\alpha_2+\alpha_4-\lambda) (\alpha'-\alpha_3-\lambda)} {\lambda (\alpha_1+\alpha_5+\alpha_3-\alpha'+\lambda)} \right\}
\end{IEEEeqnarray*}
Therefore,
\begin{IEEEeqnarray*}{rCl?C?rCl}
b'(\lambda)&<&0 &\iff& \frac{(\alpha_2+\alpha_4-\lambda) (\alpha'-\alpha_3-\lambda)} {\lambda (\alpha_1+\alpha_5+\alpha_3-\alpha'+\lambda)}&<&1\\
&&&\iff& \frac{(\alpha_2+\alpha_4)(\alpha'-\alpha_3)}{\alpha_1+\alpha_5+\alpha_2+\alpha_4}&<&\lambda
\end{IEEEeqnarray*}
which implies that $b(\lambda)$ reaches its maximum, $b$, for $\lambda=\lambda^\ast$, where $\lambda^\ast:= \frac{(\alpha_2+\alpha_4)(\alpha'-\alpha_3)}{\alpha_1+\alpha_5+\alpha_2+\alpha_4}$.

For simplicity, let $\gamma_1=\alpha_2+\alpha_4$, $\gamma_2=\alpha_1+\alpha_5$, and $\delta=\alpha'-\alpha_3$. Then, $\lambda^\ast=\frac{\gamma_1\delta}{\gamma_1+\gamma_2}$, and:
\begin{IEEEeqnarray*}{rCl}
b &=& b(\lambda^\ast)\\
&=& \frac{\psi(\gamma_1)\psi(\gamma_2)}{\psi(\lambda^\ast)\psi(\gamma_1-\lambda^\ast) \psi(\delta-\lambda^\ast)\psi(\gamma_2-(\delta-\lambda^\ast))}\\
&=& \frac{\gamma_1^{\gamma_1}}{(\lambda^\ast)^{\lambda^\ast}(\gamma_1-\lambda^\ast)^{\gamma_1-\lambda^\ast}}\\
&&\quad\times \frac{\gamma_2^{\gamma_2}}{(\delta-\lambda^\ast)^{\delta-\lambda^\ast}(\gamma_2-(\delta-\lambda^\ast))^{\gamma_2-(\delta-\lambda^\ast)}}\\
&=& \frac{\gamma_1^{\gamma_1}}{\left(\gamma_1\frac{\delta}{\gamma_1+\gamma_2}\right)^{\lambda^\ast} \left( \gamma_1\frac{\gamma_1+\gamma_2-\delta}{\gamma_1+\gamma_2} \right)^{\gamma_1-\lambda^\ast}}\\
&&\quad\times \frac{\gamma_2^{\gamma_2}}{\left(\gamma_2\frac{\delta}{\gamma_1+\gamma_2}\right)^{\delta-\lambda^\ast}\left( \gamma_2\frac{\gamma_1+\gamma_2-\delta}{\gamma_1+\gamma_2} \right)^{\gamma_2-(\delta-\lambda^\ast)}}\\
&=& \frac{\gamma_1^{\gamma_1}\gamma_2^{\gamma_2}}{\frac{\gamma_1^{\gamma_1}\delta^\delta \gamma_2^{\gamma_2} (\gamma_1+\gamma_2-\delta)^{\gamma_1+\gamma_2-\delta}}{(\gamma_1+\gamma_2)^{\gamma_1+\gamma_2}}}\\
&=& \frac{\psi(\gamma_1+\gamma_2)}{\psi(\delta)\psi(\gamma_1+\gamma_2-\delta)}\\
&=& \frac{\psi(\alpha_2+\alpha_4+\alpha_1+\alpha_5)}{\psi(\alpha'-\alpha_3)\psi(\bar\alpha-\alpha_6-\alpha')}
\end{IEEEeqnarray*}

Granted, it may be the case that $\lambda^\ast<\alpha_0-\alpha_3$, but then $b(\lambda^\ast)$ would simply be an upper bound on the actual maximum value.

\paragraph{Computing $\beta^\ast$}
Rewrite $\beta(\alpha')$ as
\begin{IEEEeqnarray*}{rCl}
\beta(\alpha') &=& P_{d^\ast}^{\alpha'}(1-p)^{\bar\alpha-\alpha'}\times b^\ast\\
&=& P_{d^\ast}^{\alpha'}(1-p)^{\bar\alpha-\alpha'} \frac{\psi(\alpha_2+\alpha_4+\alpha_1+\alpha_5)}{\psi(\alpha'-\alpha_3)\psi(\bar\alpha-\alpha_6-\alpha')}
\end{IEEEeqnarray*}

From Appendix~\ref{app:derivations}, we see that we can write:
\begin{IEEEeqnarray*}{rCl}
\sgn\{b'(\alpha')\} &=& \sgn\left\{ \log\frac{P_{d^\ast}(\bar\alpha-\alpha_6-\alpha')}{(1-p)(\alpha'-\alpha_3)} \right\}
\end{IEEEeqnarray*}
Therefore,
\begin{IEEEeqnarray*}{rCl?C?rCl}
\beta'(\alpha')&<&0 &\iff& \frac{P_{d^\ast}(\bar\alpha-\alpha_6-\alpha')}{(1-p)(\alpha'-\alpha_3)} &<& 1\\
&&&\iff& \alpha_3+\frac{P_{d^\ast}}{P_{d^\ast}+1-p}\left(\gamma_1+\gamma_2\right) &<& \alpha'
\end{IEEEeqnarray*}

Therefore, $\beta(\alpha')$ reaches its maximum for $\alpha'=\alpha^\ast$, where $\alpha^\ast:=\alpha_3+\frac{P_{d^\ast}}{P_{d^\ast}+1-p}\left(\gamma_1+\gamma_2\right)$.

However, we know that $\alpha'\ge\alpha_0$. Therefore, $\beta^\ast=\beta(\alpha^\ast)$ only if $\alpha^\ast\ge\alpha_0$. This is not always the case. When $\alpha^\ast<\alpha_0$, then the maximum over all \emph{valid} values of $\alpha'$ is reached at $\alpha_0$, since, for $\alpha'>\alpha^\ast$, $\beta(\alpha')$ is strictly decreasing. Hence, if we reintroduce the possibility that $\alpha_0$ can be infinite:
\begin{IEEEeqnarray*}{rCl}
\beta^\ast &=& \left\{\,
\begin{IEEEeqnarraybox}[][c]{l?s}
\IEEEstrut
0 & if $\alpha_0=\infty$\\
\beta(\alpha^\ast) & if $\alpha^\ast\ge\alpha_0$\\
\beta(\alpha_0) & otherwise
\IEEEstrut
\end{IEEEeqnarraybox}\right.
\end{IEEEeqnarray*}

$\alpha_0$ can take on one of two values: either $\alpha_0=\alpha_3$, or $\alpha_0=\frac{\alpha_2+\alpha_4+\alpha_3+\alpha_7}{2}$. In the former case, we will always have $\alpha^\ast\ge\alpha_3=\alpha_0$, so $\beta^\ast=\beta(\alpha^\ast)$. The latter case is the one where $\alpha^\ast<\alpha_0$ is a possibility.

Finally, we substitute the values
\begin{IEEEeqnarray*}{rCl}
\alpha^\ast&=&\alpha_3+\frac{P_{d^\ast}}{P_{d^\ast}+1-p}\left(\alpha_2+\alpha_4+\alpha_1+\alpha_5\right)\\ \alpha_0&=&\frac{\alpha_2+\alpha_4+\alpha_3+\alpha_7}{2}
\end{IEEEeqnarray*}
in the expression of $\beta(\alpha^\ast)$ and $\beta(\alpha_0)$, respectively, and get the result of Lemma~\ref{lemma:beta}, and, by extension, that of Theorem~\ref{thm:rate}.

\section{Some Supporting Mathematical Results}\label{app_Support}
This appendix gives some supporting results that are purely mathematical, and were hence left out of the previous analyses.

\subsection{Proof that $A_{m,k}\le2$}\label{app:binom-bound}

Recall that $A_{m,k}$ is defined as:
\begin{IEEEeqnarray*}{rCl}
A_{m,k} &=& \left\{\,
\begin{IEEEeqnarraybox}[][c]{l?s}
\IEEEstrut
1 & if $k=0$ or $k=m$\\
\frac{m}{k(m-k)} & otherwise
\IEEEstrut
\end{IEEEeqnarraybox}\right.
\end{IEEEeqnarray*}
for all integers $m\ge1$ and $0\le k\le m$.

For $m=1$, the only possible values of $k$ are $0$ and $1$, both of which are covered by the first case. Hence, $A_{1,k}=1\le2$ for any valid $k$.

We will therefore assume, in the following, that $m\ge2$ and $1\le k\le m-1$. We have:
\begin{IEEEeqnarray*}{C?rCl}
&\frac{m}{k(m-k)} &\le& 2\\
\iff& m &\le& 2k(m-k)\\
\iff& 2k^2-2mk+m&\le&0
\end{IEEEeqnarray*}
The roots of this quadratic function of $k$ are
\begin{IEEEeqnarray*}{rCl}
k_1 &=& \frac{1}{2}\left(m-\sqrt{m(m-2)}\right)\\
k_2 &=& \frac{1}{2}\left(m+\sqrt{m(m-2)}\right)
\end{IEEEeqnarray*}

By noting that, for $m\ge2$,
\begin{IEEEeqnarray*}{rCcCl}
\sqrt{m(m-2)} &\ge& \sqrt{(m-2)(m-2)} &=& m-2
\end{IEEEeqnarray*}
we get
\begin{IEEEeqnarray*}{rCl}
k_1 &\le& \frac{1}{2}\left(m-(m-2)\right) \le \frac{1}{2}\times(2) = 1\\
k_2 &\ge& \frac{1}{2}\left(m+(m-2)\right) \ge \frac{1}{2}(2m-2) = m-1
\end{IEEEeqnarray*}

So $k_1\le1$ and $k_2\ge m-1$. Therefore, for all values $k$ between $1$ and $(m-1)$, we have $k_1\le k\le k_2$ and thus $2k^2-2mk+m\le0$. Hence, $\frac{m}{k(m-k)}\le2$, which proves that $A_{m,k}\le2$.

\subsection{Derivatives of some useful functions}\label{app:derivations}

This section analyzes the derivatives of functions involving terms of the form $y^y$. We first deal with those functions that \emph{only} have $y^y$ terms, and then study those which we multiply by an exponential function. We use again the function $\psi(y) =y^y$, for all $y\ge0$.

\subsubsection{Functions involving only $\psi(\cdot)$}

\paragraph{Single $\psi(\cdot)$ function}
Let $f(x)\ge0$ be some differentiable function of $x$, and let $h(x)=\psi(f(x))= f(x)^{f(x)}$. To find $h'(x)$, write $h(x)=e^{f(x)\log f(x)}$, and hence:
\begin{IEEEeqnarray*}{rCl}
h'(x) &=& \left(f'(x)\log f(x) + f(x)\frac{f'(x)}{f(x)}\right)e^{f(x)\log f(x)}\\
&=& f'(x)\left[1+\log f(x)\right]h(x)
\end{IEEEeqnarray*}

\paragraph{Product of $\psi(\cdot)$ functions}
Now let
\begin{IEEEeqnarray*}{rCcCl}
h(x) &=& \prod_i h_i(x) &=& \prod_i f_i(x)^{f_i(x)}
\end{IEEEeqnarray*}

Taking the derivative, we have
\begin{IEEEeqnarray*}{rCl}
h'(x) &=& \sum_i h_i'(x)\prod_{j\not=i}h_j(x)\\
&=& \sum_i f_i'(x)\left[1+\log f_i(x)\right]h_i(x)\prod_{j\not=i}h_j(x)\\
&=& \left(\prod_ih_i(x)\right)\sum_i f_i'(x)\left[1+\log f_i(x)\right]\\
&=& h(x) \sum_i f_i'(x)\left[1+\log f_i(x)\right]
\end{IEEEeqnarray*}

\paragraph{Product and division of $\psi(\cdot)$ functions}
Finally, let
\begin{IEEEeqnarray*}{rCcCl}
h(x) &=& \frac{f(x)}{g(x)} &=& \frac{\prod_i f_i(x)^{f_i(x)}}{\prod_j g_j(x)^{g_j(x)}}
\end{IEEEeqnarray*}

Then the derivative is:
\begin{IEEEeqnarray*}{rCll}
h'(x) &=& \IEEEeqnarraymulticol{2}{l}{\frac{f'(x)g(x)-g'(x)f(x)}{g(x)^2}}\\
&=& \IEEEeqnarraymulticol{2}{l}{\frac{f(x)\sum_i f_i'(x)\left[1+\log f_i(x)\right]g(x)}{g(x)^2}}\\
&& \IEEEeqnarraymulticol{2}{l}{\:-\:\frac{g(x)\sum_j g_j'(x)\left[1+\log g_j(x)\right]f(x)}{g(x)^2}}\\
&=& \frac{f(x)}{g(x)}&\left(\sum_i f_i'(x)\left[1+\log f_i(x)\right]\right.\\
&&&\left.\quad-\:\sum_j g_j'(x)\left[1+\log g_j(x)\right]\right)\\
&=& h(x)&\left(\sum_i f_i'(x)\left[1+\log f_i(x)\right]\right.\\
&&&\left.\quad-\:\sum_j g_j'(x)\left[1+\log g_j(x)\right]\right)
\end{IEEEeqnarray*}

\paragraph{Sign of the derivative}
The main point of this analysis is to find the sign of the derivative of such functions, to find their (global) maximum.

In all the cases we will deal with, we will always have that the $f_i(x)$ and $g_j(x)$ are nonnegative functions of $x$ (actually, they take values in $[0,1]$). As a result, $h(x)\ge0$ and thus:
\begin{IEEEeqnarray*}{rCll}
\sgn\{h'(x)\} &=& \sgn&\left\{\sum_i f_i'(x)\left[1+\log f_i(x)\right]\right.\\
&&&\left.\quad\: - \sum_j g_j'(x)\left[1+\log g_j(x)\right]\right\}
\end{IEEEeqnarray*}

Moreover, all the $f_i$ and $g_j$ functions that we will deal with are \emph{linear} in $x$. Therefore, $f_i'(x),g_j'(x)\in\{-1,+1\}$. We can split them into terms with a positive derivative, and terms with a negative derivative. We will use the following notation:
\begin{IEEEeqnarray*}{rCl}
I^+ &=& \{ i : f_i'(x)=+1 \}\\
I^- &=& \{ i : f_i'(x)=-1 \}\\
J^+ &=& \{ j : g_j'(x)=+1 \}\\
J^- &=& \{ j : g_j'(x)=-1 \}
\end{IEEEeqnarray*}

Therefore,
\begin{IEEEeqnarray*}{rCll}
\IEEEeqnarraymulticol{4}{l}{\sgn\{h'(x)\}}\\
\qquad &=& \sgn&\left\{ \sum_i f_i'(x)\left[1+\log f_i(x)\right]\right.\\
&&&\left.\quad\: - \sum_j g_j'(x)\left[1+\log g_j(x)\right] \right\}\\
&=& \sgn&\left\{ \sum_{i^+\in I^+}[1+\log f_{i^+}] - \sum_{i^-\in I^-}[1+\log f_{i^-}]\right.\\
&&&\left.\quad\: - \sum_{j^+\in J^+}[1+\log g_{j^+}] + \sum_{j^-\in J^-}[1+\log g_{j^-}] \right\}\\
&=& \sgn&\left\{ |I^+|+|J^-|-|I^-|-|J^+|\right.\\
&&&\left.\quad\: + \log\frac{\Bigl( \prod_{i^+} f_{i^+}\Bigr) \Bigl(\prod_{j^-}g_{j^-}\Bigr)} {\Bigl(\prod_{j^+}g_{j^+} \Bigr) \Bigl(\prod_{i^-}f_{i^-}\Bigr)} \right\}
\end{IEEEeqnarray*}

Moreover, since all of our cases will have $|I^+|+|J^-|=|I^-|+|J^+|$, then
\begin{IEEEeqnarray*}{rCl}
\sgn\{h'(x)\} &=& \sgn\left\{ \log\frac{\Bigl( \prod_{i^+} f_{i^+}(x)\Bigr) \Bigl(\prod_{j^-}g_{j^-}(x)\Bigr)} {\Bigl(\prod_{j^+}g_{j^+}(x) \Bigr) \Bigl(\prod_{i^-}f_{i^-}(x)\Bigr)} \right\}
\end{IEEEeqnarray*}

By expressing the sign of $h'(x)$, we can now easily find the value of $x$ that maximizes the value of functions of the same form as $h(x)$.

\subsubsection{Functions involving $\psi(\cdot)$ terms as well as an exponential}

Recall that, if $q(x)=a^xb^{m-x}$, for some constants $a$, $b$ and $m$, then
\begin{IEEEeqnarray*}{rCl}
q'(x) &=& a^xb^{m-x}\log\frac{a}{b}\\
&=& q(x)\log\frac{a}{b}
\end{IEEEeqnarray*}

Now, if we have $h_2(x)=q(x)h_1(x)$, with:
\begin{IEEEeqnarray*}{rCl}
h_1(x) &=& \frac{\prod_i f_i(x)^{f_i(x)}}{\prod_j g_j(x)^{g_j(x)}}
\end{IEEEeqnarray*}
and with $h_1$ having $|I^+|+|J^-|=|I^-|+|J^+|$, then,
\begin{IEEEeqnarray*}{rCll}
h_2'(x) &=& \IEEEeqnarraymulticol{2}{l}{q'(x)h_1(x)+q(x)h_1'(x)}\\
&=& \IEEEeqnarraymulticol{2}{l}{q(x)\left(\log\frac{a}{b}\right)h_1(x)}\\
&&\:+q(x)h_1(x)&\left( \sum_i f_i'(x)\left[1+\log f_i(x)\right]\right.\\
&&&\left.\quad\: - \sum_j g_j'(x)\left[1+\log g_j(x)\right] \right)\\
&=&\IEEEeqnarraymulticol{2}{l}{h_2(x) \log \frac{a\Bigl( \prod_{i^+} f_{i^+}(x)\Bigr) \Bigl(\prod_{j^-}g_{j^-}(x)\Bigr)} {b\Bigl(\prod_{j^+}g_{j^+}(x) \Bigr) \Bigl(\prod_{i^-}f_{i^-}(x)\Bigr)}}
\end{IEEEeqnarray*}

Again, looking at the sign, if we assume $a,b\in[0,1]$ (as is the case in our problem), then $h_2(x)\ge0$ and
\begin{IEEEeqnarray*}{rCl}
\sgn\{h_2'(x)\} &=& \sgn\left\{ \log \frac{a\Bigl( \prod_{i^+} f_{i^+}(x)\Bigr) \Bigl(\prod_{j^-}g_{j^-}(x)\Bigr)} {b\Bigl(\prod_{j^+}g_{j^+}(x) \Bigr) \Bigl(\prod_{i^-}f_{i^-}(x)\Bigr)} \right\}
\end{IEEEeqnarray*}

\end{document}